\documentclass{jpsj-suppl}
\usepackage{times}
\usepackage{color}
\usepackage{bm}
\usepackage{amsmath, amsfonts}
\title{Study of interface phenomena in a topological-insulator/Mott-insulator heterostructure}

\author{$^{1}$Suguru Ueda\thanks{E-mail address: ueda@scphys.kyoto-u.ac.jp}, $^{1}$Norio Kawakami and $^{2}$Manfred Sigrist }
\inst{$^{1}$ Department of Physics, Kyoto University, Kyoto 606-8502, Japan \\
 $^{2}$ Theoretische Physik, ETH Z\"{u}rich, CH-8093 Z\"{u}rich
}
\abst{ 
We theoretically investigate a two-dimensional heterostructure composed of a topological insulator (TI) and a Mott insulator (MI), and clarify what kind of electronic states can be realized at the interface. By using inhomogeneous dynamical mean-field theory, we confirm that the topological edge state penetrating into the MI region induces a heavy-fermion like mid-gap state. We further elucidate the nature of the spatially-modulated quasi-particle weight of the mid-gap state, and discuss the effects of local correlation in the TI region. The optical conductivity and the Drude weight are also computed with changing the electron tunneling near the interface.
}

\kword{topological insulator, heterostructure, electron correlation, \ldots}

\begin{document}
\maketitle

\section{Introduction}

Since the experimental investigation of HgTe/CdTe quantum wells~\cite{bernevig2006quantum,konig2007quantum} and some bismuth compounds, such as Bi$_2$Se$_3$ and Bi$_{1-x}$Sb$_x$~\cite{zhang2009topological,xia2009observation,hsieh2008topological,hsieh2009observation}, topological insulators (TI) have attracted considerable attention as a platform for novel condensed matter physics. The most distinctive signature of TIs is the existence of the gapless edge or surface state, which is intrinsically linked to the topological nature of electronic bulk spectrum. Thereby, the edge state of the TIs is robust against the perturbation without breaking time-reversal symmetry.

In the study of the TI, the interest has recently also turned towards heterostructures involving TI, and extensive theoretical and experimental 
activities focused on the exploration of Majorana fermions potentially realized at interfaces~\cite{PhysRevLett.100.096407,PhysRevLett.102.216404,das2012zero} and on possible applications in spintronics devices~\cite{qi2008fractional,pesin2012spintronics}. There is also the important question on how strong correlation effects manifest themselves via topological edge states in TI/strongly-correlated-electron-systems (SCES) heterostructures, because most previous works have studied the interface between the TI and non (or weakly)-interacting electrons. Actually, it has been reported that correlation effects lead to exotic and various interface phenomena in the heterostructures of the SCES~\cite{okamoto2004electronic,PhysRevLett.101.066802} : for instance, two-dimensional (2D) metallic state and superconductivity are observed at the interface of trivial band-insulators~\cite{ohtomo2002artificial,reyren2007superconducting}. Furthermore, it is known that the Mott physics even supports novel topological states, such as a topological-quantum phase without gapless edge states~\cite{PhysRevB.83.205122}. Thereby, it is interesting to shed light on the interface electronic properties in the TI/SCES.

In this study, we investigate 2D heterostructures composed of a paramagnetic Mott insulator (MI) and a TI by using a simple microscopic model. The many-body effects are treated by means of the inhomogeneous dynamical-mean-field theory, and we discuss the interplay of the local Coulomb interaction and the topological edge states penetrating into the MI. In our previous study, we have elucidated that the penetration of the topological edge state drives the renormalized mid-gap state inside the MI region~\cite{PhysRevB.87.161108}. We clarify, in this work, the spatial modulation of the mid-gap state, particularly focusing on its quasi-particle weight with varying the local interaction. We also discuss how the electron hopping around the interface affects the optical conductivity and the Drude weight.

\section{Model and Formalism}

\subsection{ Model of a MI/TI heterostructure }

%
\begin{figure}[t]
\centering
\includegraphics[width=0.4\linewidth]{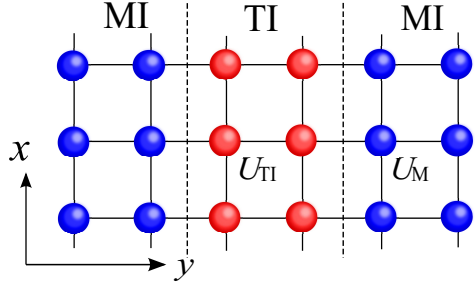}
\caption{ \label{fig:model} Sketch of the present heterostructure described by the tight-binding model. Inside the MI (TI) region, electrons move on a lattice of blue (red) sites, interacting with Coulomb repulsion $U_{\text M}$ ($U_{\text {TI}}$). In this study, the width of the MI and TI regions are set 10 and 20 unit cells, respectively. }
\end{figure}
%

In order to discuss generic features of the MI/TI interface, we consider 2D heterostructure with a simple-cubic lattice as a minimal microscopic model. Figure~\ref{fig:model} displays a schematic figure of this heterostructure. The corresponding Hamiltonian is given by $H = H_{\text{TI}} +H_{\text{MI}} +H_{\text{V}}$ with,
%
%
%
\begin{eqnarray}
H_{\text {TI}} &=& \sum _{i,\sigma,\alpha} \epsilon _{\alpha} \hat {n}_{i \sigma \alpha}
  +\sum _{\substack {\langle i,j \rangle,\\ \sigma, \alpha, \beta}}
   \hat {a}^{\dagger}_{i \sigma \alpha} \left [ \hat {t}_{\sigma}(\delta) \right ]_{\alpha \beta} \hat {a}_{j \sigma \beta}
  +U_{\text{TI}}\sum_{i \alpha} \hat{n}_{i \uparrow  \alpha}\hat{n}_{i \downarrow  \alpha} \label{eq:bhz} \\
H_{\text {MI}} &=& \sum _{\langle i,j \rangle, \sigma, \alpha} t_{\alpha}
						\hat {c}^{\dagger}_{i\sigma \alpha}\hat{c}_{j\sigma \alpha} +U_{\text{M}}\sum _{i \alpha} \hat{n}_{i \uparrow \alpha}\hat{n}_{i \downarrow \alpha}, \label{eq:hub} \\
H_{\text {V} } &=& \sum _{\langle i,j \rangle, \sigma, \alpha} V_{\alpha} \left (
						\hat{c}^{\dagger}_{i \sigma \alpha}\hat{a}_{j \sigma \alpha} +\hat{a}^{\dagger}_{i \sigma \alpha}\hat{c}_{j \sigma \alpha} \right ) \label{eq:hybridization}.
\end{eqnarray}
%
Here, an annihilation operator $\hat c_{i \sigma \alpha}$ ($\hat a_{i \sigma \alpha}$) acts on the orbital-$\alpha=1,2$ in the MI (TI) region, and the matrix $\hat t(\delta)$ with $\delta = \pm x(\pm y)$ in Eq.~(\ref{eq:bhz}) is defined by,
%
\begin{eqnarray}
\hat {t}_{\sigma}(\pm x) = 
	\begin{pmatrix}
	t_1 \:\:\: \pm i \sigma t_{so} \\
	\pm i \sigma t_{so} \:\:\: t_2
	\end{pmatrix}, \:\:\:
\hat {t}_{\sigma}(\pm y) =
	\begin{pmatrix}
	t_1 \:\:\: \pm t_{so} \\
	\mp t_{so} \:\:\: t_2
	\end{pmatrix}.
\end{eqnarray}
%
In these expressions, $t_{so}$ is interband hybridization, $U_{\text {M}}$ and $U_{\text {TI}}$ are local Coulomb interaction parameters and $V_{\alpha}$ is the amplitude of the electron hopping between the TI and the MI. The MI and the TI are described by a two band Hubbard model $H_{\text{M}}$ and a generalized Bernevig-Hughes-Zhang (BHZ) model $H_{\text{TI}}$~\cite{PhysRevB.85.165138}, respectively, and $H_{\text{V}}$ is the tunneling matrix at the interface. 
Throughout this paper, we assume $V_1=-V_2=-V$ for simplicity, and choose the parameters as $t_1=-t_2=-t$, $\epsilon _1 = -\epsilon _2 =-t$ and $t_{so}=0.25t$. Our study is restricted to the particle-hole symmetric and non-magnetic case at zero temperature (thus, the spin index is dropped). In addition, we set the width of the MI and the TI regions are 10 and 20 unit cells, respectively, and choose the $y$-axis perpendicular to the interface.

\subsection{ Inhomogeneous dynamical mean-field approach }

In our numerical analysis, periodic (open) boundary conditions areadopted for $x$($y$)-direction, and the Hubbard repulsion is treated with the inhomogeneous dynamical mean-field theory (IDMFT)~\cite{PhysRevLett.101.066802, PhysRevB.59.2549, PhysRevB.70.241104,PhysRevB.79.045130}. In the framework of the IDMFT, the lattice problem is mapped onto an impurity problem for each $y$, and then the self-energy is obtained as a function of $y$. In addition, taking into account the result of the previous study~\cite{PhysRevB.85.165138}, we set the inter-band component of the self-energy $\Sigma _{12,21}(\omega)$ to zero. With this assumption, the self-consistent equation in IDMFT can be reduced to the following simple one: 
%
\begin{eqnarray}
\mathcal {G}_{0\alpha}^{-1}(y;\omega ) &=& 
	\left [ \int \frac{d k_x}{2\pi} G_{\alpha}(y;k_x, \omega ) \right ]^{-1}
	+\Sigma_{\alpha}(y; \omega ), \\
G_{\alpha}(y;k_x, \omega ) &=& \left [ \,(\omega +\mu)\mathbb{I} -\hat H_0(k_x) -\hat {\Sigma }(\omega) \right ] ^{-1}_{yy,\alpha}.
\end{eqnarray}
%
Here, $\mu$ is the chemical potential, $\hat {\mathcal {G}}_{0\alpha}$ is the cavity Green's function, and $\hat H_0(k_x)$ is the Fourier transform of the non-interacting part of the present Hamiltonian. 

To solve the impurity problem, we employ the exact diagonalization methods (ED)~\cite{PhysRevLett.72.1545}, which is suitably applied to the IDMFT analysis~\cite{PhysRevB.79.045130}. In the ED calculation, the impurity bath sites are approximated by a finite number of bath sites 
and the corresponding bath parameters are determined by minimization of the following function:
%
\begin{equation}
\chi (y) = \sum _{\omega _n} | \, {\mathcal {G}_{0}^{\text {Ns}}}(y; i\omega _n) -{ \mathcal{G} }_{0}(y, y; i\omega _n)|^2,
\end{equation}
%
where, $\omega_n$ is the discretized Matsubara frequency $\omega_n = (2n + 1)\pi/\tilde{ \beta }$ and ${\mathcal {G}_{0}^{\text {Ns}}}$ is the non-interacting Green's function for the system described by $N_s$ bath sites. In the numerics, the Lanzcos algorithm is employed to solve the local Green's function with $N_s = 7$ and the fictitious temperature $\tilde{ \beta } = 200$.

\section{Results and discussion}

We present the computed results in this section. In the calculation, the MI (TI) layers are resided on $y=-1,-2,-3,...$ ($y=0,1,2,...$) so that $y=-1$ ($y=0$) labels the edge of the MI (TI) regions. In the bulk limit, the two-band Hubbard model~(\ref{eq:hub}) has a Mott transition as $U_{\text{M}}$ exceeds a critical value $U_c \sim 13.2t$. Hence, unless otherwise mentioned, we choose $U_{\text{M}}=13.3t$ which is slightly larger than $U_c$. Additionally, following the treatment in Ref.~\cite{PhysRevB.85.165138}, we use the terms such as renormalized electrons, quasi-particle states, etc., which are characteristic of Fermi-liquid systems.

\subsection{ Spatial modulation of the spectral weight}
%
\begin{figure}[t]
\centering
\includegraphics[width=0.5\linewidth]{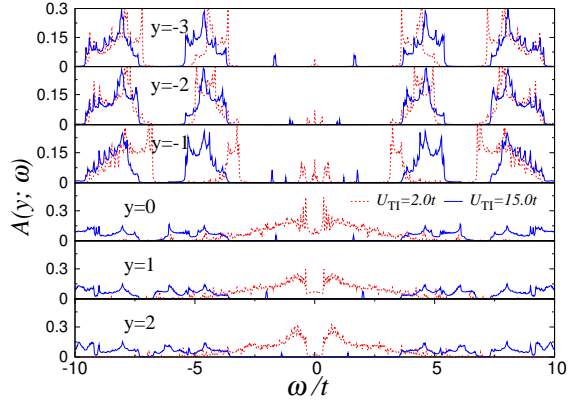}
\caption{ \label{fig:dos} Layer resolved spectral density around the interface with $U_{\text M}=13.3t$ and $V=t$. The solid (dotted) curve shows the results for $U_{\text{TI}}=15.0t$ $(2.0t)$. The MI and the TI layers are arranged in $y=-1,-2,-3,...$ and $y=0,1,2,...$, respectively. }
\end{figure}
%

In the present system, we confirm that the energy-spectrum at the edge of the TI ($y=0$) shows a single Dirac cone dispersion at $k_x=\pi$, which is a distinctive feature of the interface between topologically trivial and non-trivial materials. In our previous work~\cite{PhysRevB.87.161108}, it is elucidated that the TI edge states spread into the MI region, and forms the heavy-fermion-like mid-gap state within the Hubbard gap. Figure ~\ref{fig:dos} displays the corresponding layer-resolved spectral function $A(y; \omega) = -(1/\pi) \sum _{\alpha, k_x} \text{Im} G_{\alpha }(y, y; k_x, \omega +i\delta)$ around the interface (see dotted line). It is also pointed out that the renormalized mid-gap state shows the helical Dirac-like dispersion as a remnant of the penetrating edge state. 
%
\begin{figure}[b]
\hspace {0.02\linewidth}
\includegraphics[width=0.45\linewidth]{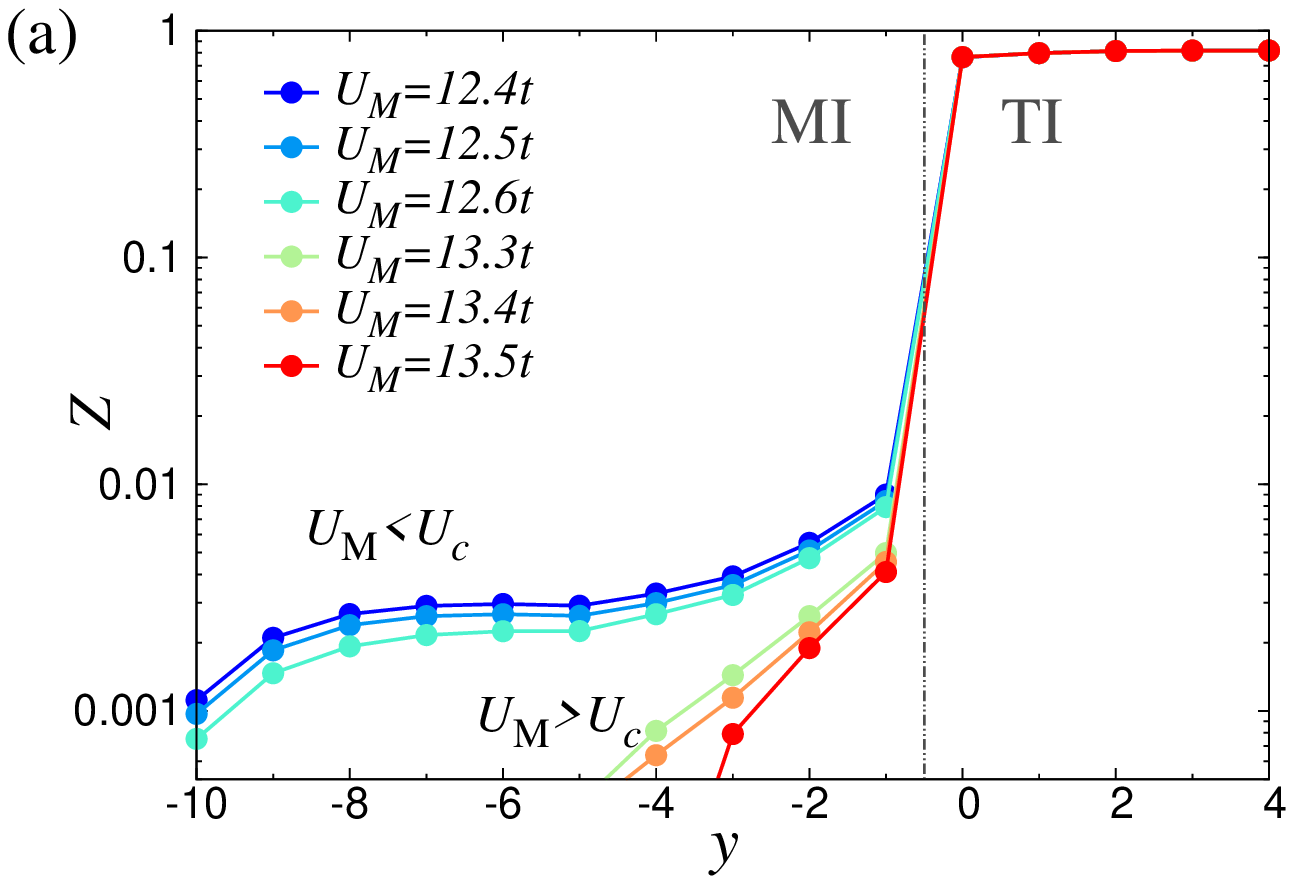}
\hspace {0.02\linewidth}
\includegraphics[width=0.46\linewidth]{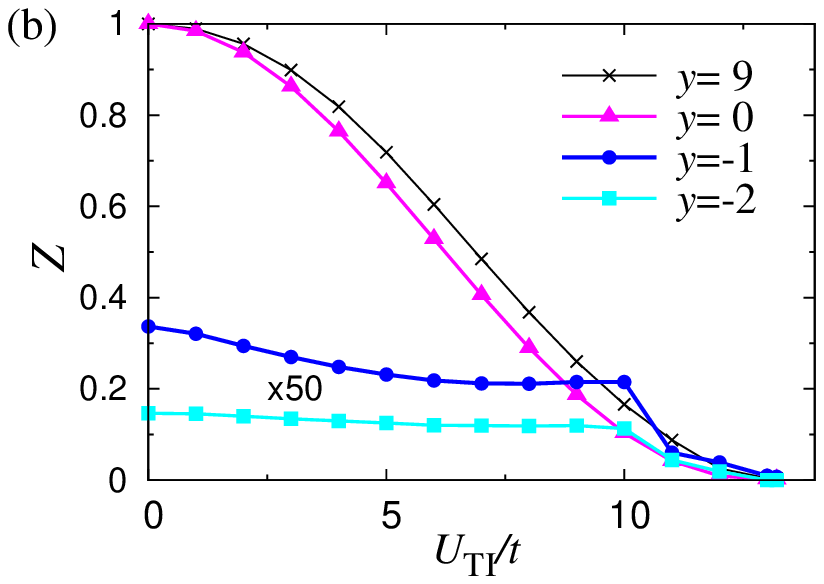}
\caption{ \label{fig:renorm_z} (a) The plot of the  quasi-particle weight as a function of $y$ with $V=t$, $U_{\text {TI}}=4t$ and several values of $U_{\text M}$; $U_{\text M}=12.4t, 12.5t, 12.6t$ for $U_{\text M} < U_c$ and  $U_{\text M}=13.3t, 13.4t, 13.5t$ for $U_{\text M} > U_c$. (b) The plot of the quasi-particle weight $Z(y)$ as a function of $U_{\text {TI}}$ with $U_{\text M} = 13.3t$ and $V=t$. The triangle, circle and square symbols correspond to $Z(y)$ at $y = 0, -1$ and $-2$, respectively; $Z(y = -1)$ and $Z(y = -2)$ are enlarged 50 times. For the comparison, $Z(y)$ at $y=9$ (the center of the TI) is also indicated. }
\end{figure}
%

Toward further understanding of the nature of the renormalized mid-gap state, we start with discussion on the layer dependence of the quasi-particle weight $Z(y)$ defined by,
%
\begin{equation}
Z(y) = \left [ 1- \frac{ \partial \text{Im} \Sigma(y; i\omega ) }{\partial \omega } \right ]^{-1} \simeq \left [ 1- \frac{ \text{Im} \Sigma(y; i\omega _0) }{\omega _0} \right ]^{-1},
\end{equation}
%
where $\omega _0$ is the lowest Matsubara frequency~\cite{PhysRevLett.102.206407}. Figure ~\ref{fig:renorm_z} (a) represents the plot of $Z(y)$ as a function of $y$ with varying the on-site Coulomb interaction $U_{\text {M}}$ below and above the critical value $U_c$. We note that the sudden suppression of $Z(y)$ for $U_{\text M}<U_c$ around $y=-10$ can be understood in terms of the surface effect, which enhances the electron correlation due to the reduced coordination number. As can be seen in Fig.~\ref{fig:renorm_z} (a), $Z(y)$ shows the stronger $U_{\text M}$-dependence inside the MI rather than the interface. For $U_{\text {M}}<U_c$, $Z(y)$ forms the plateau away from the interface ($y<-5$), and recovers its bulk property. On the contrary, $Z(y)$ shows an exponential decay in the MI region for $U_{\text {M}}>U_c$. Importantly, such layer- and $U_{\text {M}}$-dependences of $Z(y)$ are very similar to that of the well-studied metal/Mott heterostructure~\cite{PhysRevLett.101.066802,PhysRevB.85.115134}. This implies that the physical aspects of the mid-gap state are mainly dominated by the local nature of the MI region, and thus the characteristic helical energy spectrum would disappear rapidly if we are slightly away from the interface, in comparison with $U_{\text {M}}<U_c$.

Next let us turn to the effects of the onsite Coulomb interaction in the TI region. In Fig.~\ref{fig:renorm_z} (b), $Z(y)$ is plotted as a function of $U_{\text{TI}}$ for $y=-2, -1, 0, 9$. As can be seen, $Z(y)$ in the TI gradually approaches to zero. On the other hand, $Z(y)$ at $y=-2,-1$ shows weak $U_{\text {TI}}$-dependence for $U_{\text {TI}}<10t$, while its magnitude is abruptly reduced in the vicinity of the Mott transition point. This implies that the nature of the mid-gap state is closely linked to the renormalization of the edge state in the strongly correlated regime. We finally comment on the Mott physics of the present system. As reported in many previous studies, the local electron correlation competes with and even destroys the topological phases. This is also true for the renormalized mid-gap state, because it stems from the emergent edge state. Actually, once a Mott transition occurs in the TI with increasing $U_{ \text{TI}}$, the quasi-particle weight simultaneously goes to zero in all layers. The solid lines in Fig.~\ref{fig:dos} correspond to the spectral function in the Mott-insulating state for $U_{\text {TI}} = 15t$. We note that, although the TI is sandwiched by the MIs, the nature of transition is consistent with the previous study for the TI thin-film~\cite {PhysRevB.85.115134}, and no signature of the layer selective transition~\cite{PhysRevB.83.205122} is observed within our single-site DMFT analysis.

\subsection{ Drude weight of electrons at the interface}

%
\begin{figure}[b]
\hspace {0.02\linewidth}
\includegraphics[width=0.445\linewidth]{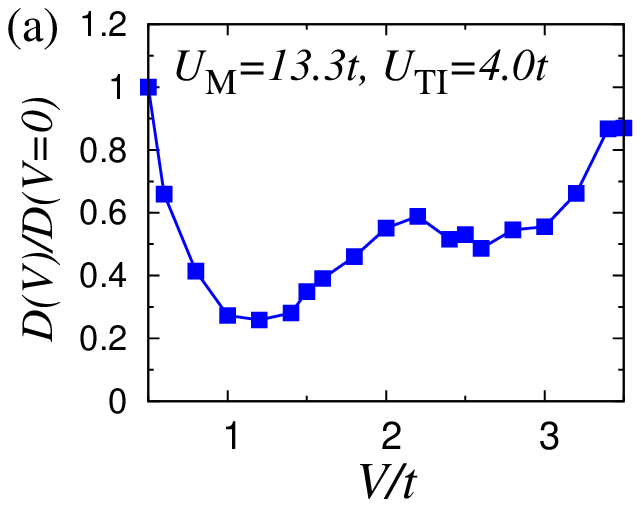}
\hspace {0.02\linewidth}
\includegraphics[width=0.465\linewidth]{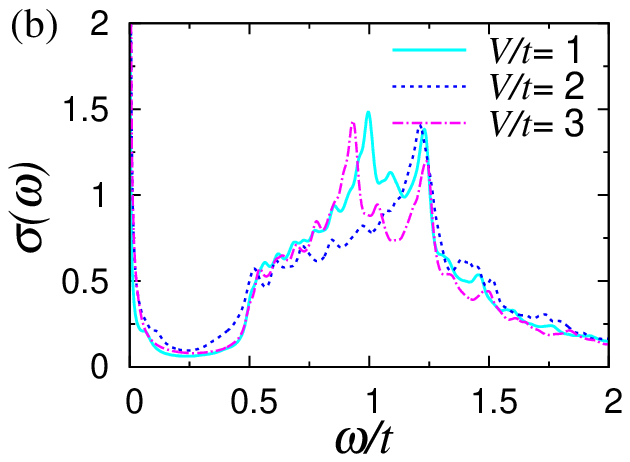}
\caption{ \label{fig:drude} (a) $V$-dependent Drude weight $D(V)$ with $U_{\text M}=13.3t$ and $U_{\text {TI}}=4.0t$. $D(V)$ is normalized by using the value at $V=0$. (b) Corresponding optical conductivity for $V=t, 2t, 3t$, which corresponds to the solid, dotted and dashed lines, respectively. }
\end{figure}
%
A key quantity realizing the mid-gap state is the electron tunneling $V$ between the MI and the TI. In this subsection, we study how the increasing $V$ modifies the transport properties with electronic field along the $x$-direction. In Fig.~\ref{fig:drude}, we plot the Drude weight and the corresponding optical conductivity $\sigma (\omega)$ with several values of $V$, computed from the effective low-energy Hamiltonian, $\hat Z^{1/2} [ \hat H_0(k_x) -\mu \mathbb{I} +\text {Re} \hat {\Sigma}(0) ] \hat Z^{1/2}$~\cite{PhysRevB.70.241104}. The quasi-particle contribution to the conduction is well captured by the Drude weight $D$. It is found that, with increasing $V$, the curve of the Drude weight changes its slope in the vicinity of $V \sim 1.5t$ as shown in Fig.~\ref{fig:drude} (a). In the present system, the transport properties are mainly dominated by the contribution of the topological edge state. Therefore, the monotonic suppression of $D$ for small $V$ would be explained by the (nearly) flat spectrum of the mid-gap state, because it reduces the Fermi velocity of the edge state thorough the interlayer hybridization $V$. However, since further increasing $V$ strongly disturbs the Dirac dispersion, the situation becomes more complicated. In order to understand the upturn of $D$, it is necessary to consider the dimerization nature around the interface. As discussed in Ref.~\cite{PhysRevB.87.161108}, large $V$ forms the gapped structure between $y=-1$ and $y=0$, and associated with this, the Dirac dispersion is reconstructed at $y=1$ in the limit of $V \rightarrow \infty$. Thereby, the magnitude of $D$ approaches to that at $V\sim 0$ for large $V$, as can be seen in Fig.~\ref{fig:drude} (a).

These features are also confirmed in the obtained optical conductivity in Fig.~\ref{fig:drude} (b). Here, $\sigma (\omega)$ mainly consists of two parts: the Drude peak located at $\omega=0$ and the broad hump for $0.5t \lesssim \omega \lesssim 1.5t$ reflecting the bulk spectrum of the TI. We note that the latter one also shows the distinctive $V$-dependence. While the peak structure around $\omega \sim t$ is washed out for $V \sim 2.0t$, its magnitude is recovered with further increasing $V$. As discussed above, this is understood by the contribution of the edge spectrum: the edge state has the large spectral weight around $\omega \sim \pm 0.5t$ (see Fig.~\ref{fig:dos}), and actually its magnitude shows the similar $V$-dependence against the increase of $V$.

\section{Summary}

We studied the 2D model heterostructure consisting of the TI and the MI, and investigated the proximity effect by using IDMFT. In this system, the topological edge states penetrate into the MI region, inducing strongly renormalized mid-gap states with a helical energy spectrum, whose quasiparticle weight decays exponentially for strong onsite repulsion. 
With tuning the interaction in the TI, a Mott transition occurs in the TI, such that both the mid-gap and the topological edge states simultaneously disappear. The optical conductivity and the Drude weight as a function of the electron tunneling $V$ between the TI and the MI visualizes well the variation of interface electronic properties.

This work was supported by the Grant-in-Aid for the Global COE Program ''The Next Generation of Physics, Spun from Universality and Emergence'' from MEXT of Japan. N. K. is supported by KAKENHI (Nos.22103005,25400366) and JSPS through its FIRST Program and S.U. by a JSPS Fellowship for Young Scientists. We also acknowledge support from the Swiss Nationalfonds and the Pauli Centre of ETH Zurich. 

\appendix

\end{document}